# Y a-t-il élimination d'*Eupelmus orientalis* Crawford par *Eupelmus vuilleti* Crawford (Hymenoptera : Eupelmidae) des systèmes de stockage du niébé (*Vigna unguiculata* Walp) ?

## A. Ndoutoume-Ndong[(1)], D. Rojas-Rousse[(2)]

[(1)]Ecole Normale Supérieure de Libreville, B.P. 17009 Libreville- Gabon- Tel. (241) 26 16 54 fax. (241) 73 31 61- email : augustendoutoume@caramail.com

[(2)]Institut de Recherche sur la Biologie de l'Insecte, UPRESA-CNRS 6035, Faculté des Sciences, Parc de Grandmont, 37200 Tours – France.

Titre courant : Elimination d'*Eupelmus orientalis* Crawford ?

**Résumé**

Le niébé est une légumineuse alimentaire cultivée en Afrique tropicale pour ses graines riches en protéines. Le principal problème que pose sa production est la conservation des récoltes. Dans les champs comme dans les stocks, les graines sont détruites par des insectes ravageurs (les bruches). Ces bruches sont toujours associées à plusieurs espèces d'hyménoptères entomophages. Quatre espèces d'entomophages ont été recensées : une espèce oophage (*Uscana. lariophaga* Stephan), et trois espèces larvo-nymphales (*Dinarmus basalis* Rondoni, Pteromalidae; *Eupelmus vuilleti* Crawford et *Eupelmus orientalis* Crawford, Eupelmidae). Ces deux Eupelmidae et *D. basalis* sont des ectoparasitoïdes solitaires. Le suivi des populations montre qu'au début du stockage, *E. orientalis* est l'espèce la plus abondante (72 %) alors qu'E. *vuilleti D. basalis* représentent respectivement 12% et 16 % des hyménoptères. Durant le stockage, les effectifs de la population d'*E. orientalis* diminuent progressivement et elle disparaît complètement en moins de deux mois après le début du stockage. *E. vuilleti* devient progressivement majoritaire aux dépens de *D. basalis* dont les effectifs régressent jusqu'à moins de 10 % des parasitoïdes émergeants. *E. vuilleti* adopte un comportement ovicide et larvicide vis-à-vis de *D. basalis*, ce qui explique la régression de ses effectifs des systèmes de stockage. Si le comportement de cet Eupelmidae est une constante, cela pourrait aussi expliquer la disparition d'*E. orientalis* des stocks. Or si cette espèce se maintenait dans les stocks, elle serait un bon auxiliaire de lutte contre les bruches compte tenu de ses capacités parasitaires. Cette étude montre que le comportement ovicide et larvicide d'*E. vuilleti* ne s'exprime pas vis-à-vis d'*E. orientalis*. Lorsque les femelles ne disposent exclusivement que des hôtes déjà parasités par *E. orientalis*, elles préfèrent s'abstenir de pondre. La disparition d'*E. orientalis* ne saurait donc s'expliquer par la présence d'*E. vuilleti*.

**Abstract**

Niébé is a food leguminous plant cultivated in tropical Africa for its seeds rich in proteins. The main problem setted by its production is the conservation of harvests. In the fields as in the stocks, the seeds are destroyed by pests (bruchids). These bruchids are always associated with several entomophagous species of hymenoptera. Four entomophagous species were listed : an egg parasitoid (*U lariophaga Stephan*), and three solitary larval and pupal ectoparasitoids (*D. Basalis Rondoni*, Pteromalidae; *E. vuilleti Crawford* and *E. orientalis Crawford*, Eupelmidae). The survey of the populations shows that at the beginning of storage, *E orientalis* is the most abundant specie (72 %) whereas *E. vuilleti* and *D. Basalis* respectively represent 12 % and 16 % of the hymenoptera. During storage, the *E orientalis* population decreases gradually and it disappears completely in less than two months after the beginning of storage. *E. Vuilleti population* becomes gradually more important than *D. basalis* population which regress until less than 10 % of the emerging parasitoids. *E vuilleti* adopts ovicide and larvicide behaviour against *D. Basalis*. This behaviour explains its population regression inside granaries. If the aggressive behaviour of this Eupelmidae is a constant, that could also explain the disappearance of *E orientalis*. However if this species is maintained in stocks, it would be an effective control agent of bruchids according to their parasitic capacities. This study shows that ovicide and larvicide behaviour of *E vuilleti* is not expressed against *E orientalis*. When the females have exclusively the hosts already parasitized by *E orientalis*, they do not lay eggs. The disappearance of *E orientalis* could not thus be explained by the presence of *E. vuilleti*.

# Introduction

En Afrique tropicale, le niébé (*Vigna unguiculata* Walp) est une légumineuse alimentaire largement cultivée. Pour les pays de la zone sahélienne, tel que le Niger, la production de niébé est estimée à 300 000 tonnes par an (Alzouma 1981). Les graines de cette légumineuse, constituent la source de protéines la moins onéreuse pour la plupart des populations africaines. En effet, les graines de niébé contiennent la plupart des acides aminés nécessaires à l'alimentation humaine, à l'exception des acides aminés soufrés (Smart 1964). C'est donc un aliment de haute valeur nutritive qui pourrait aider les populations locales dans leur effort vers l'autosuffisance alimentaire. Dans les sols riches et irrigués, les rendements sont élevés et sa commercialisation représente une source de revenus importante. Au delà de son intérêt alimentaire, le niébé revêt donc un intérêt économique certain.

Cependant, la production du niébé est limitée par les pertes occasionnées par des insectes Coléoptères Bruchidae qui rendent très difficile sa conservation après la récolte. En effet, les stades larvaires de ces insectes ravageurs se développent à l'intérieur des graines et consomment les réserves contenues dans les cotylédons. Un taux d'infestation initial des graines de 10 % par des larves de ces bruches (*Callosobruchus maculatus* Fabricius) suffit pour détruire en quelques mois 60 à 70 % de la récolte du niébé (Gauthier 1996). En outre, au cours de leur développement, les larves de bruches éliminent l'azote sous forme d'acide urique toxique qui s'accumule à l'intérieur des graines, ce qui rend le niébé parasité impropre à la consommation.

Dans les champs comme dans les systèmes de stockage du niébé, les études ont montré que les populations de bruches sont toujours associées à plusieurs espèces d'hyménoptères entomophages (Germain *et al.* 1987). Dans la région de Niamey par exemple, quatre espèces de parasitoïdes ont été recensées : une espèce oophage (*Uscana lariophaga* Stephan, Trichogrammatidae), et trois espèces (*Dinarmus basalis* Rondoni, Pteromalidae; *Eupelmus vuilleti* Crawford et *Eupelmus orientalis* Crawford, Eupelmidae) qui sont des ectoparasitoïdes solitaires des larves et des nymphes. Le suivi des populations dans les greniers traditionnels montre qu'au début du stockage, *E. orientalis* Crawford est l'espèce la plus abondante et représente 72 % des émergences d'hyménoptères à cette période alors qu'*E. vuilleti* Crawford et *D. basalis* Rondoni sont moins abondants et représentent respectivement 12 % et 16 % des hyménoptères émergeant des graines (Monge & Huignard, 1991). Durant le stockage, les effectifs de la population d'*E. orientalis* Crawford diminuent progressivement et disparaissent complètement 45 jours environ après le début du stockage (Monge & Huignard, 1991). Lammers et Van Huis (1989), puis Monge et Huignard (1991) montrent par contre que les deux autres espèces de parasitoïdes larvaires se maintiennent dans les greniers tout au long de la période de stockage. *E. vuilleti* Crawford devient progressivement majoritaire aux dépens de *D. basalis* Rondoni dont les effectifs régressent jusqu'à moins de 10 % des parasitoïdes émergeants.

Les travaux réalisés par Leveque *et al.* (1993), Van Alebeek *et al.* (1993) montrent qu'il y a une compétition interspécifique entre *E. vuilleti* Crawford et *D. basalis* Rondoni. Ces auteurs démontrent qu'*E. vuilleti* Crawford adopte un comportement ovicide (multiparasitisme) et larvicide (hyperparasitisme) vis-à-vis de *D. basalis* Rondoni. Selon Monge *et al.* (1995), le comportement agressif d'*E. vuilleti* Crawford est reponsable de la régression des effectifs de *D. basalis* Rondoni des systèmes de stockage. Si l'agressivité de cet Eupelmidae est une constante, cela pourrait aussi expliquer la disparition d'*E. orientalis* Crawford des greniers. Or le maintien de cette espèce dans les stocks serait intéressant dans la mise au point de programme de lutte biologique contre les bruches (Doury 1995). En effet, *E. vuilleti* Crawford qui se maintient dans les stocks a malheureusement une plus faible capacité reproductrice. A chaque génération, *E. vuilleti* Crawford n'arrive à parasiter que 10 à 30 % des hôtes disponibles (Huignard & Monge, 1993) alors qu'*E. orientalis* Crawford parasite à chaque génération 55 à 71 % des hôtes disponibles (Ndoutoume *et al.* 2000). Dans ce travail, nous avons cherché à montrer si *E. vuilleti* manifeste un comportement agressif qui pourrait expliquer la disparition d'*E. orientalis*.

## *Matériel et Méthode*

### Conditions d'élevage au laboratoire

La souche de *Callosobruchus maculatus* utilisée comme hôte primaire pour cette étude a été fondée depuis 14 mois à partir des centaines d'individus émergeant des gousses de niébé récoltées dans des champs au Burkina-Faso. Elle a été maintenue au laboratoire sur des graines de niébé de la variété California Black Eyes, dans les conditions climatiques suivantes : 40°C-25°C, L : D 12 : 12, 30 ± 10% h. r. (40°C le jour et 25 ° C la nuit , avec 12 h de jour et 12 h d'obscurité, l'humidité relative étant de 30 ± 10%). Ces conditions ont été choisies afin de limiter les attaques et l'installation des acariens dans les élevages. Sous ces conditions *C. maculatus* est polyvoltine, à raison d'une génération tous les 23 à 25 jours. L'élevage de masse a lieu dans des boîtes en plexiglas (17,5 x 11,5 x 3 cm) où plusieurs couples de bruches de forme "non voilière" (incapable de voler) sont placés en présence de graines de niébé pendant 24 à 48 heures. A l'issue de ce temps, les parents sont retirés et les graines sont conservées dans l'étuve jusqu'à l'émergence des descendants qui sont aussi des individus de forme non voilière.

Les parasitoïdes utilisés au cours de ce travail proviennent d'une souche fondée à partir d'individus émergeant de gousses de niébé récoltées dans des champs de la région de Niamey (13° N) depuis 9 mois. Dès leur émergence, les adultes d'*E. orientalis* et d'*E. vuilleti* sont placés dans des cages d'élevage différentes. Chacune de ces cages en plexiglas ($40 \times 30 \times 25$ cm) contient des graines de niébé infestées de bruches, déposées dans des couvercles de boîtes de Pétri, et de l'eau saccharosée pour l'alimentation. Tous les trois jours, les graines de niébé sont renouvelées et celles qui sont retirées sont conservées dans les boîtes de Petri jusqu'à l'émergence des parasitoïdes adultes qui sont à leur tour remis dans les cages d'élevage dans les mêmes conditions que les bruches. Pour éviter le phénomène de multiparasitisme, les deux espèces de parasitoïdes sont élevées dans des pièces différentes. Ce sont les larves L4, prénymphes et nymphes de *C. maculatus* Fabricius qui sont utilisées comme hôtes ; ces stades correspondent respectivement à 15, 16-17, et 18-19 jours après la ponte.

*E. orientalis* Crawford et *E. vuilleti* Crawford sont des ectoparasitoïdes solitaires des larves et nymphes de bruches. L'œuf est déposé sur le tégument de *C. maculatus* Fabricius et le développement pré et post-embryonnaire a lieu à l'intérieur de la loge creusée par ce dernier. Ces Eupelmidae ont cinq stades larvaires pourvus de mandibules qui jouent un rôle important lors des combats larvaires qui ont lieu en situation de super et de multiparasitisme (Delanoue & Arambourg 1965; Leveque *et al.* 1993; Cortesero 1994; Doury 1995). Le temps de développement dure en moyenne 20 jours chez *E. orientalis* Crawford et 23 jours chez *E. vuilleti* Crawford (Ndoutoume *et al.* 2000) Après la mue imaginale, l'adulte émerge de la graine par un petit opercule circulaire découpé dans le tégument de la graine. Les mâles émergent deux à trois jours avant les femelles et l'accouplement peut avoir lieu aussitôt après l'émergence.

Dès l'émergence, pour chaque espèce, les femelles sont isolées et réparties en lots de cinq individus avec 5 mâles âgés de deux à trois jours. Après l'accouplement surveillé, chaque femelle est transférée dans une petite boîte cylindrique en plexiglas (3,5 cm de hauteur et 8 cm de diamètre). Chaque femelle, nourrie d'eau saccharosée, est placée en présence de graines de niébé renfermant des bruches (*C. maculatus* Fabricius). Les hôtes offerts sont des œufs, des prénymphes ou des nymphes parce que les parasitoïdes y pondent préférentiellement et cela évite de biaiser la sex-ratio en faveur des mâles (Terrasse 1986; Terrasse *et al.* 1996).

**Paramètres mesurés**

L'activité parasitaire des femelles *E. vuilleti* Crawford a été évaluée par le décompte du nombre d'œufs pondus ($1^{er}$ lot) et du nombre de descendants émergés ($2^{ième}$ lot). Ces paramètres ont été analysés sur des hôtes sains (nymphe de *C. maculatus* Fabricius), parasités par des oeufs, des larves et des nymphes d'*E. orientalis* Crawford.

Pour analyser l'activité parasitaire des femelles *E. vuilleti* Crawford et les descendances observées dans les différentes situations expérimentales, nous avons utilisé le test de Chi-carré ($\chi 2$).

***Etude de l'activité parasitaire des femelles E. vuilleti Crawford en situation de choix de la qualité des hôtes***

Cette expérience doit permettre de différencier le comportement de parasitisme des femelles *E. vuilleti* Crawford en présence des hôtes sains, parasités par elles-mêmes et par les femelles *E. orientalis* Crawford. Cette situation est plus proche de celle qu'on retrouve dans les greniers traditionnels africains où le niébé est stocké. Pour cela, une femelle âgée de 5 jours est placée en présence de 8 hôtes durant 4 jours successifs :

   2 hôtes contenant chacun une larve de dernier stade ($L_5$) *E. orientalis* Crawford.

   2 hôtes contenant chacun une larve de dernier stade ($L_5$) *E. vuilleti* Crawford.

4 hôtes sains (nymphes de *C. maculatus*) : c'est le lot témoin.

### *Etude de l'activité parasitaire des femelles* **E. vuilleti** *Crawford* *sur des hôtes préalablement parasités par* **E. orientalis** *Crawford*

Ce test permet d'étudier le comportement des femelles *E. vuilleti* Crawford lorsqu'elles se retrouve en situation extrême, c'est-à-dire uniquement en présence des hôtes parasités par *E. orientalis* Crawford. Vont-elles s'abstenir de pondre ou pas lorsqu'elles n'ont pas de choix ? C'est une situation qui n'est pas très réaliste car, dans les greniers, il y tous les types d'hôtes (hôtes parasités et hôtes sains).

Trois tests ont été conduits en parallèle. Dans chacun d'eux, l'activité individuelle d'une femelle est analysée en présence de 5 hôtes préalablement parasités par *E. orientalis* Crawford. Pour augmenter la pression parasitaire de la femelle sur les hôtes, les mêmes hôtes sont laissés 4 jours successifs et changés 5 fois au cours de la vie de la femelle. Cela signifie qu'une femelle a reçu 25 hôtes (5 hôtes x 5) sur 20 jours (1 lot de 5 hôtes tous les 4 jours). Ces 20 jours couvrent les 2/3 de la durée moyenne de vie d'une femelle *E. vuilleti* Crawford.

**Premier test : hôtes parasités porteurs d'œufs d'***E. orientalis* Crawford. Le comportement de parasitisme de 40 femelles a été analysé (20 femelles pour la répartition de la ponte, 20 femelles pour la production des parasitoïdes adultes). Dans chaque lot, 500 hôtes ont été présentés aux femelles.

**Deuxième test : hôtes parasités porteurs d'une larve du cinquième stade d'***E. orientalis* Crawford. 25 femelles ont été testées pour la répartition de la ponte et 25 autres pour la descendance engendrée ; dans chaque situation 625 hôtes ont été offerts aux femelles.

**Troisième test : hôtes parasités porteurs d'une nymphe d'***E. orientalis* Crawford. Le comportement de parasitisme de 40 femelles a été analysé (20 pour l'analyse de la répartition de la ponte et 20 pour l'analyse de la descendance engendrée). Dans chaque lot 500 hôtes ont été présentés aux femelles.

## *Résultats*

**Activité parasitaire des femelles *E. vuilleti* Crawford en situation de choix de la qualité des hôtes**

**Sur larves L5-hôtes.** L'ouverture de la graine rendant l'hôte accessible à l'expérimentateur permet : 1) de vérifier l'état réel de l'hôte offert aux femelles, 2) de vérifier le parasitisme secondaire de ces hôtes. Dans nos conditions expérimentales, sur les 2 x 80 hôtes parasités au préalable par les 2 Eupelmides, 72 étaient réellement porteurs d'une L5 d'*E. vuilleti* Crawford et, 64 d'une L5 d'*E. orientalis* Crawford (tab. 1).

Il y a évitement du parasitisme secondaire (c'est à dire de l'hyperparasitisme) sur la presque totalité des hôtes car les L5 *E.v*-hôtes ont été évités pour la plupart (71 rejetés sur les 72 porteurs d'une L5 soit 95% de rejet) tout comme ceux porteurs d'une L5 *E.o.* (57 évités sur les 64 offerts soit 89% de rejet) (tab. 1). Cela signifie que les femelles d'*E. vuilleti* Crawford n'ont hyperparasité qu'un hôte porteur d'une larve d'un congénère et **7** porteurs d'une L5 de l'espèce voisine. L'hyperparasitisme de sa propre espèce ou d'*E. orientalis* Crawford semble donc être évité par les femelles d'*E. vuilleti* Crawford.

**Sur les hôtes sains**. Sur les 160 hôtes sains offerts, 140 d'entre eux ont été parasités soient 90,3% des hôtes offerts (tab. 1). On a dénombré en moyenne 2 oeufs ou/et larves néonates par hôte.

**Pour ce qui concerne le suivi des lots d'hôtes destinés à l'étude de la descendance**

**Sur larves L5-hôtes.** Après l'émergence des parasitoïdes (30 à 35 jours après le début de l'expérimentation : délai établi en fonction de la durée moyenne du développement des parasitoïdes qui est de 20 jours), l'ouverture des graines d'où aucun adulte n'a émergé (bruches ou parasitoïdes), permet de vérifier a posteriori l'état réel des graines qui ont été présentées aux femelles. Dans la série des hôtes (larves L5 d'*E. vuilleti*), seulement 54 graines contenaient des L5-hôtes car les 26 autres renfermaient une bruche morte. La presque totalité des graines renfermant une L5-hôte a permis le développement d'un parasitoïde adulte : 50 ont donné naissance à 1 adulte *E. vuilleti* Crawford (tab. 1). La production en adultes *E. vuilleti* est donc de 93%. Dans la série des hôtes (larves L5 d'*E. orientalis*), la production en adultes *E. orientalis* Crawford est aussi élevée : 94% (tab. 1). Ces valeurs sont à rapprocher de celles observées dans le rejet des L5-hôtes évalué par la répartition de la ponte :

cas L5*Ev*-hôtes : rejet 95% et production d'adultes 93%.

cas L5*Eo*-hôtes : rejet 89% et production d'adultes 94%.

Dans chacun des cas, les valeurs proches des pourcentages révèlent que ce sont bien les larves L5 développées en parasitoïdes primaires qui ont bien donné des adultes. Cela signifie que les femelles *E. vuilleti* Crawford ont bien évité, dans nos deux tests, d'hyperparasiter leurs congénères ou l'espèce voisine.

**Sur les hôtes sains.** La descendance observée est de 124 adultes (86 femelles + 38 mâles), soit 77,50% des hôtes fournis. Ces données renforcent celles déduites de l'analyse de la répartition de la ponte.

**Activité parasitaire des femelles *E. vuilleti* Crawford sur des hôtes préalablement parasités par *E. orientalis* Crawford**

La distribution des oeufs (sur la cible alimentaire ou en dehors), pondus par les femelles en présence des hôtes offerts tous les quatre jours, montre que ce sont les hôtes porteurs d'œufs d'*E. orientalis* Crawford qui reçoivent le plus d'œufs (tab. 2). Ces hôtes sont multiparasités puisque l'hôte de base (larve de bruche) est encore 'intacte' car sa consommation par la larve néonate issue de l'œuf pondu 36 heures auparavant (celui d'*E. orientalis* Crawford) n'a pas débuté. Compte tenu du laps de temps entre les deux pontes, la larve d'*E. orientalis* Crawford éclos en premier et tue l'œuf pondu par *E. vuilleti* Crawford. Malgré tout, le niveau de ponte reste toujours bas en présence des hôtes préalablement parasités par *E. orientalis* Crawford.

En absence de choix des hôtes, les femelles *E. vuilleti* Crawford multiparasitent faiblement les hôtes porteurs d'œufs d'*E. orientalis* Crawford alors qu'elles rejettent la majorité des hôtes porteurs de stades larvaires plus avancés L5 ou nymphe d'*E. orientalis* Crawford (tab. 2).

**Analyse de la descendance**

Les distributions des adultes, établies en fonction des trois types d'hôtes offerts, montrent leur opposition entre la production en *E. vuilleti* Crawford et *E. orientalis* Crawford. Dans les trois situations étudiées, de tous les hôtes offerts, des *E. orientalis* Crawford émergent en grande majorité (tab. 3). Cela signifie que la majorité des hôtes offerts ont échappé au multiparasitisme ou à l'hyperparasitisme de la part des femelles d'*E. vuilleti* Crawford. Le nombre d'*E. orientalis* Crawford émergés des hôtes offerts est significativement plus élevé par rapport à celui des *E. vuilleti* Crawford quel que soit le stade de développement de l'hôte (tab. 3).

## *Discussion*

Au cours de cette étude, l'activité parasitaire individuelle des femelles *E. vuilleti* Crawford a été analysée dans deux situations. La première, qui représente l'expérience témoin, a permis de voir qu'*E. vuilleti* Crawford évite les hôtes déjà parasités (par ses congénères ou par les femelles d'*E. orientalis*) et parasite les hôtes sains lorsque les deux types d'hôtes sont disponibles. Cela montre que dans nos conditions expérimentales, les femelles *E. vuilleti* Crawford parasitent de préférence les hôtes sains (90,3% contre 1,4% et 10,9%). La présence d'hôtes parasités ne paraît pas préjudiciable à l'activité de parasitisme des femelles *E. vuilleti* Crawford sur les hôtes sains. Le rejet des hôtes parasités (L5-hôtes), correspond donc bien à un choix délibéré des femelles et non à une perturbation de leur comportement de parasitisme.

Dans la deuxième situation, pour augmenter la pression parasitaire, *E. vuilleti* Crawford n'a eu que des hôtes *E. orientalis* Crawford à disposition. Les femelles n'avaient pas le choix de la qualité de leurs hôtes. Les mêmes hôtes sont restés 4 jours en présence des mêmes femelles qui n'avaient pour seul choix que pondre ou ne pas pondre. Devant une telle contrainte, les femelles *E. vuilleti* Crawford ont très faiblement parasité les hôtes contenant des *E. orientalis* Crawford (105 sur 500 hôtes-œufs offerts, 58 sur 625 hôtes-L5 d'*E. orientalis* et 40 sur 500 hôtes-L5 *E. vuilleti*). Les femelles *E. vuilleti* Crawford s'abstiennent donc de parasiter les hôtes préalablement infestés par *E. orientalis* Crawford. Ce comportement d'évitement se confirme par le nombre important d'œufs, de larves et nymphes *E. orientalis* Crawford qui atteignent le stade adulte en présence des femelles d'*E. vuilleti* Crawford: en moyenne 75 % des *E. orientalis* arrivent au stade adulte. Ce comportement est totalement opposé à celui que cette espèce manifeste vis-à-vis de *D. basalis* Rondoni. Dans la même situation, *E. vuilleti* Crawford parasite jusqu'à 78 % d'hôtes contenant des oeufs de *D. basalis* Rondoni (Van Alebeek *et al.* 1993).

Dans les systèmes de stockage, les femelles *E. vuilleti* Crawford sont en présence d'une gamme d'hôtes sains et parasités. En effet, en situation de choix, les femelles *E. vuilleti* Crawford parasitent de préférence les hôtes sains (90,3 % des hôtes sains sont parasités contre 10,9 % des hôtes contenant des *E. orientalis*). Le comportement agressif manifesté par *E. vuilleti* Crawford en présence de *D. basalis* Rondoni (Alebeek *et al.* 1993 ; Leveque *et al.* 1993), ne s'exprime plus vis-à-vis d'*E. orientalis* Crawford même si les femelles disposent exclusivement d'hôtes déjà parasités. *E. vuilleti* Crawford est donc capable non seulement de faire la différence entre les hôtes sains et les hôtes parasités, mais aussi entre les hôtes parasités par des espèces différentes. Dans nos conditions expérimentales, le comportement agressif connu chez cet Eupelmidae disparaît vis-à-vis d'une espèce phylogénétiquement plus proche.

De tels phénomènes de reconnaissance interspécifique ont déjà été mis en évidence entre espèces du genre *Aphytis* (Rosen & DeBach 1979), entre les espèces du genre *Asobara* (Vet *et al.* 1984), entre espèces du genre *Aphidius* (McBrien & Mackauer 1990 ; 1991), puis entre *Aphidius ervi* Haliday (Braconidae) et *Aphelinus asychis* Walker (Aphelinidae) (Bai & Mackauer 1991). Pour Van Alphen et Visser (1990), le multiparasitisme est un phénomène commun dans la nature et la discrimination interspécifique n'existe qu'entre espèces très proches. Cela est confirmé par la plasticité comportementale d'*E. vuilleti* qui accepte le multiparasitisme vis-à-vis d'une espèce phylogénétiquement éloignée *(D. basalis* : Pteromalidae) alors qu'il évite complètement le multiparasitisme vis-à-vis de l'espèce sœur (*E. orientalis*). La plasticité comportementale des femelles *E. vuilleti* ne permet pas d'expliquer la disparition d'*E. orientalis* des systèmes de stockage.

Des études réalisées par Rasplus (1988), rapportent qu'*E. orientalis* présente une distribution géographique très large. cette espèce a été décrite d'Inde mais sa présence en Afrique du Sud avait déjà été mentionnée par Nikol'skaya (1963). Elle est mentionnée aussi en Afrique Centrale et de l'Ouest où elle parasite 20 espèces d'hôtes différents (18 espèces de Bruchidae et 2 espèces d'Apionidae). Cette large gamme trophique fait qu'elle est rencontrée dans la nature presque tout au long de l'année (11 mois sur 12). En Afrique de l'Ouest, cet Eupelmidae ne montre pas de préférence nette pour un milieu car il se rencontre aussi bien en forêt galerie qu'en savane, sur des herbacées ou sur des arborés (Rasplus 1988). Les stocks constituent pour elle un milieu confiné et que sa disparition est due à la colonisation de nouveaux habitats suite à une évasion des greniers.

## Littérature citée

**Tableau 1** : Activité parasitaire des femelles *Eupelmus vuilleti* Crawford en présence d'hôtes sains (nymphe de *C. maculatus*) et d'hôtes déjà parasités (le dénominateur représente les effectifs réels d'hôtes porteurs de L5 *E.vuilleti* Crawford ou *E.orientalis* Crawford dénombrés à l'ouverture des graines)

|  | L5 Ev - hôtes | L5 Eo - hôtes | Hôtes Sains |
|---|---|---|---|
| **Fréquence des hôtes avec des oeufs** $\frac{\text{Hôtes hyperparasités ou parasités}}{\text{Hôtes disponibles}}$ | $\frac{1}{72} = 0{,}014$ | $\frac{7}{64} = 0{,}109$ | $\frac{140}{155} = 0{,}903$ |
| **Fréquence des hôtes donnant des adultes** (1 adulte = 1 hôte) $\frac{\text{Parasitoïdes adultes}}{\text{Hôtes disponibles}}$ | $\frac{50}{54} = 0{,}926$ | $\frac{60}{64} = 0{,}937$ | $\frac{124}{160} = 0{,}775$ |

**Tableau 2** : Activité parasitaire des femelles *Eupelmus vuilleti* Crawford sur des hôtes préalablement parasités par *Eupelmus orientalis* Crawford en situation de non choix (n = nombre d'hôtes offerts)

|  | Hôtes-oeufs d'E. *orientalis* | Hôtes-L5 d'E. *orientalis* | Hôtes-nymphes d'E. *orientalis* | Valeur de $\chi^2$ |
|---|---|---|---|---|
| **Proportion d'hôtes parasités** | 21 % <br> n = 500 | 9,3 % <br> n = 625 | 8 % <br> n = 500 | 48,12 <br> ( **p < 0,001**) |

**Tableau 3** : Descendance observée sur des hôtes préalablement parasités par *Eupelmus orientalis* Crawford en présence des femelles *Eupelmus vuilleti* Crawford (n = total d'hôtes offerts).

|  | **Hôtes-oeufs d'E. orientalis** | **Hôtes-L5 d'E. orientalis** | **Hôtes-nymphes de E orientalis** |
|---|---|---|---|
| **- Proportion d'hôtes ayant donné des *E. vuilleti*** | 7 %<br>n = 500 | 6 %<br>n = 625 | 3 %<br>n = 500 |
| **- Proportion d'hôtes ayant donné des *E. orientalis*** | 74 %<br>n = 500 | 78 %<br>n = 625 | 73 %<br>n = 500 |
| **- Valeur de $\chi^2$** | 92,9<br>( $p < 0,001$) | 103,3<br>( $p < 0,001$) | 103,9<br>( $p < 0,001$) |